# Streamlining Compliance and Risk Management with RegTech Solutions

Chintamani Bagwe

Citibank, USA

## Abstract

*RegTech is a rapidly rising financial services sector focused on using cutting-edge technology to improve the process of regulatory compliance. RegTech solutions are characterized by numerous features and benefits that can considerably contribute to helping organizations operate effectively in the increasingly regulated environment, when it comes to compliance and risk management. This paper sheds light on why RegTech will be one of the most promising markets, driven by the rising cost of compliance and the growing reliance on technology in crisis management. Moreover, this paper will examine the advantages of using such solutions to strike a balance between compliance and operational efficiencies. This paper will deepen the understanding of regulatory compliance, introduce RegTech, and examine the benefits of using these solutions to achieve compliance.*

## 1. Introduction to RegTech Solutions

The field of financial services is constantly changing, and every day regulators worldwide disrupt the traditional compliance and risk management status quo by demanding adherence to new rules and requirements threatened by the state of the global financial market, technology, and new developments. Financial institutions are forced not only to comply with the constantly changing requirements most efficiently but also to maximize the efficiency of their business operations and strengthen competitiveness. Adhering to the RegTech – or Regulatory Technology – concept offers a clearer answer to these questions by leveraging new technologies implementation.

RegTech is a rapidly growing area that includes the creative use of technology to ensure adherence to established legal and regulatory norms. It implements several technological solutions in its framework, such as artificial intelligence (AI), machine learning (ML), data analysis, cloud computing, and blockchain, to improve the efficiency of the compliance environment. Used in conjunction, these technologies are designed to help businesses respond better to changing regulations, facilitating compliance and effectively managing risk.

A significant advantage [1] of RegTech is that its technologies are intended to streamline several routine compliance procedures. Processes traditionally performed manually, which can consume a lot of time and subject to normal human performance-error conditions, can be significantly improved with automation. Modern RegTech platforms use unique algorithms and machine-learning models to identify authentic activities [2] by tracking transactions and spotting irregularities in real-time. This constant monitoring includes not only a high rate of monitoring but also makes the performance of such essential checks simpler for the user.

Moreover, RegTech solutions are equipped with sophisticated data analytics tools, offering profound insights into compliance patterns and potential threats. RegTech tools can analyse ample data from multiple sources, detecting patterns while predicting potential compliance breaches. This allows financial institutions to proactively manage risks and react accordingly to regulatory changes, saving hundreds of billions in penalties and market value.

Outwardly, RegTech's applications in cross-industry are not limited to compliance monitoring. Such tools are capable of carrying full-scale risk assessments and simplify regulatory reporting and data protection. For example, RegTech platforms in the cloud provide accessible and flexible services connecting seamlessly with the organization's mainframe, ensuring a smooth exchange of data and hassle-free operations. Also, in terms of blockchain shaping, RegTech enhances network transparency and traceability.

Besides that, RegTech offers financial institutions a competitive advantage in the marketplace. Much of the routine compliance work is automated with the aid of RegTech, freeing up time and resources to concentrate on business goals. By doing so, they are developing a climate of innovation that reduces threats to competition.

To conclude, RegTech will be one of the most promising markets, driven by the rising cost of compliance and the growing reliance on technology in crisis management. However, complex the arena might seem, RegTech helps firms run a safe and peaceful business. This paper will deepen the understanding of regulatory compliance, introduce RegTech, and examine the advantages of using such solutions to strike a balance between compliance and operational efficiencies.

## 2. UNDERSTANDING REGULATORY COMPLIANCE

Regulatory compliance can be defined as the adherence to laws, regulations, and standards provided by regulatory authorities in a given industry. From the definition, one may deduce that compliance is an essential component of a business unit, especially in a highly policed sector such as finance. It is hard to focus more on regulatory compliance, mainly because it strives to set the industry standards [3] by which all employees abide by safeguarding the consumer's well-being and safeguarding transparency.

Among the reasons why regulatory compliance is imperative include: it helps an organization be sensitive to the legal context and avoid the repercussions of the law. Besides, it helps structure the industry in terms of good competition and avoiding rogue firms that may quickly lift themselves. Finally, sensitivity to consumer needs and ensuring they are well protected from various dangers that hurt them.

Regulatory compliance is essential to the financial industry and organizations find it challenging [4] to achieve. The financial industry is among the most sensitive sectors and requires exclusive control to curb illegal practices such as money laundering, misuse of customers' information, and mangling untouched such as saving and microcredits.

The adoption of RegTech solutions has brought a tremendous change into the compliance field for financial institutions. RegTech is short for Regulatory Technology, which means adopting cutting-edge technology such as artificial intelligence, machine learning, and data analytics to automate compliance task, real-time monitoring of transactions, and identification of potential risk. Remarkably, RegTech solutions reduce compliance processes significantly by automation. This is achieved through reducing working due to automation, improving accuracy and giving

International Journal of Computer Science & Information Technology (IJCSIT) Vol 16, No 3, June 2024

compliance teams actionable insight. Therefore, financial institutions have escaped compliance problems such as information overload and staffing.

Moreover, the use of RegTech solutions avails a great opportunity for financial institutions to efficiently manage their compliance responsibilities [5]. Availing RegTech solutions offers financial institutions a competitive edge to keep pace at the same time reducing risk. The effect of regulatory change is unavoidable; therefore, financial institutions are in the best position to handle it effectively. Compliance and risk management solutions provide an excellent opportunity to prepare innovate, stay within the law, and act as best practice: Such as ever-changing financial compliance landscape, utilizing RegTech is more of philosophy than a logical strategy. Therefore, an overview of RegTech is essential since it impacts greatly on the compliance and role of the regulatory body.

## 3. OVERVIEW OF REGTECH

Regulatory Technology, well-known as RegTech, refers to a rapidly growing financial services sector focused on using advanced technology to improve the process of regulatory compliance. With the regulatory environment becoming more complicated and fast-paced, RegTech solutions have become critical for the ability of financial institutions to ensure compliance as well as optimize their operations.

RegTech solutions utilize a variety of advanced technologies, including artificial intelligence, machine learning, data analytics, cloud computing, and blockchain, among others, to automate and simplify compliance-related tasks. Thus, these tools assist organizations in the efficient management of the rapidly expanding intricacy of regulatory demands. For instance, AI and ML enable the evaluation of vast data streams to reveal patterns, anomalies, and predict occurrences of future non-compliance. Cloud computing solutions are scalable and capable of being integrated into existing systems without disintegration to guarantee seamless, real-time data exchange. The following are some general areas of application for RegTech:

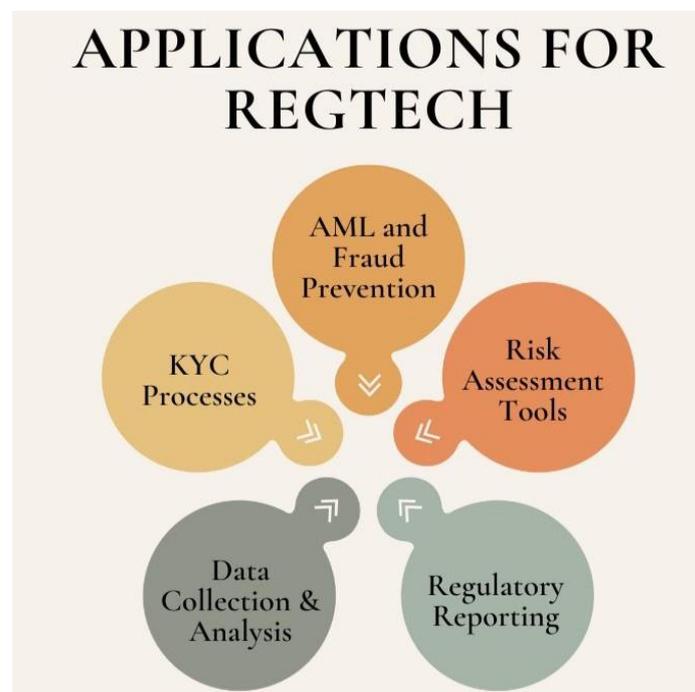

Figure 1. RegTech Applications

19

International Journal of Computer Science & Information Technology (IJCSIT) Vol 16, No 3, June 2024

- AML and Fraud Prevention involve tools [6] offering real-time transaction monitoring and risk assessment capabilities to identify and prevent money laundering and fraud by rapidly identifying and flagging dubious behaviours.
- Risk Assessment tools use data analytics [7] and ML to conduct broad assessments and provide a view of their most serious danger zones, supporting proactive danger management.
- Regulatory Reporting involves automated instruments and innovative software reporting technology approach the job of assembling and submitting regulatory reports, allowing organizations to quickly and effectively submit reports and decrease their danger of failing to adhere to regulatory mandates.
- Data Collection and Analysis enables [8] companies to manage its information quickly and efficiently, using both structured and unstructured data to provide value without being overwhelmed.
- KYC Processes is another area, since they allow automation across identity verification, customer due diligence, and ongoing monitoring, ensuring compliance with customer-related regulations

RegTech has shown to be increasingly valuable because of the need for financial service providers to become more efficient and lower their expenses automatically. Traditional procedures may be time-consuming, susceptible to human error, and might delay achieving regulatory mandates because they necessitate extensive manual investigation. RegTech may assist alleviate the burden on compliance teams by significantly automating the process.
Aside from aiding in the automation of labour-intensive work, RegTech will offer systemic benefits to businesses. The efficiency gains associated with the automated system serve as a significant competitive edge in themselves. Businesses with RegTech will be capable of reacting to regulatory updates rapidly. RegTech will enable firms to handle emergent risks quickly and more confidently by flagging them instantly as they emerge.

In conclusion, RegTech is transforming the financial business by introducing new possibilities in terms of risk management and compliance. Since regulatory restrictions become more complex and demanding on firms, RegTech offers a strong structure for addressing those needs. The adoption of RegTech is critical for any financial institution wishing to grow concurrently with the financial sector and its own turnover.

## 4. BALANCING COMPLIANCE AND EFFICIENCY

The advent of RegTech solutions in the financial industry has changed the way to maintain a balance between compliance and operational efficiency. Compliance requirements become more complicated, as modern businesses cannot afford to waste resources on ensuring compliance with current regulations. However, overemphasis on compliance may prevent modern organizations from the flexibility and innovation driving the markets.

RegTech solutions introduce a new approach to ensuring compliance by implementing machine learning and artificial intelligence to comply easily with regulations. Precise and fast data collection and analysis, monitoring transactions in real-time, and assessing risks are some of the features provided by RegTech. The main challenges towards this balance are the growing complexities of regulatory requirements, a dynamic market landscape, the data collection and analysis burden, constant vigilance from the regulators, and numerous risks.

## 4.1. Challenges in Balancing Compliance and Efficiency

Striking a balance between compliance and efficiency may be challenging for companies. Compliance standards are becoming more intricate as a result of the growing number of regulatory requirements and the vibrant financial environment. A careful examination of facts, regulatory oversight, and constantly burgeoning potential risks contribute to this complexity.
Companies with highly regulated operations require the fine line among compliance and extremity. Firms in highly controlled businesses can use this next step RegTech solutions to increase automation procedures effortlessly commissioning the newest advanced. Companies can deliver more than just compliance reduction coaching by integrating RegTech; this strategy can provide them a significant advantage.

## 5. KEY FEATURES AND BENEFITS OF REGTECH SOLUTIONS

Overall, RegTech solutions are characterized by numerous features and benefits that can considerably contribute to helping organizations operate effectively in the increasingly regulated environment when it comes to compliance and risk management. These characteristics include the use of innovative technology, as well as its ability to automate, accelerate, and ensure compliance with regulatory aspects and requirements. Some of the mentioned features and benefits are as follows:

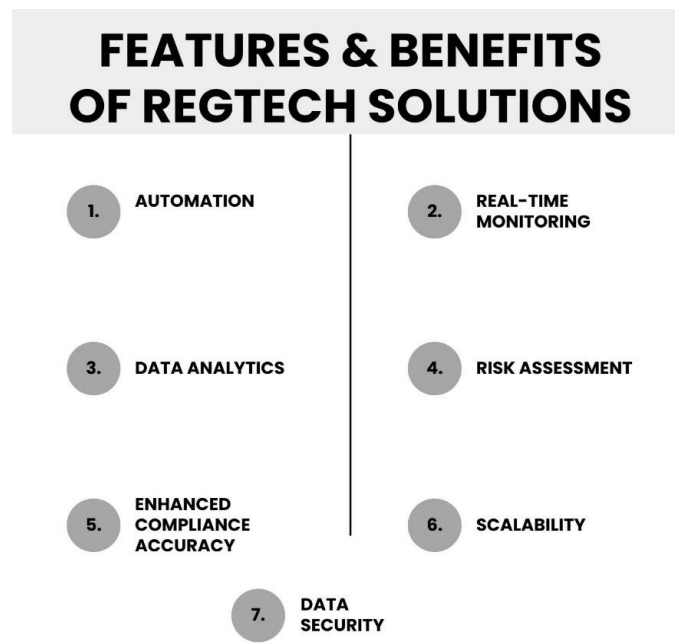

Figure 2. RegTech Benefits

## 5.1. Automation

RegTech solutions automate most compliance activities thus eliminating the occurrence of human errors. They reduce the time taken for numerous inauspicious compliance activities such as data collection and analysis, data dissemination, reporting among others.

## 5.2. Real-Time Monitoring

RegTech solutions give an organization a real-time view of all transactions or activities happening within the organization. As a result, the organization is capable of identifying and responding to compliance violations in real-time hence averting any associated penalties.

## 5.3. Data Analytics

RegTech solutions utilize big data analytics thus able to analyse a vast volume of data and provide valuable insights driving compliance and risks. Organizations that utilize RegTech can view data in colour, hence a better position to notice compliance violations and other non-conforming activities.

## 5.4. Risk Assessment

RegTech augments an organization's ability to conduct a risk assessment by using complex models and algorithms thus effective identification, assessment, and mitigation. With RegTech, an organization can avoid heavy fines hence upholding its good standing and that of its stakeholders.

## 5.5. Enhanced Compliance Accuracy

RegTech aids an organization in remaining accurate and compliant with numerous standards. They have standard processes and controls thus an employee operating it will be less compromised to forget to adhere to compliance and standards.

## 5.6. Scalability

RegTech facilitates change and can effortlessly grow with one's business as new dynamics such as regulatory come up.

## 5.7. Data Security

Finally, RegTech solutions offer exceptional data security. They protect sensitive information from unauthorized access, breaches, and cyber threats. Businesses can rely on their own security measures, such as encryptions and secure data storage, to keep the data safe and ensure compliance with data protection regulations.

To sum up, RegTech solutions provide many valuable features and benefits that transform compliance and risk management experience in the financial industry. With automation tools, real-time monitoring, data analytics, and risk assessment capacities, businesses can streamline their compliance processes, move through the maze of regulations, and gain a competitive advantage in the changing nature of regulations.

## 6. USE CASES OF REGTECH IN FINANCIAL INSTITUTIONS

More financial institutions are increasingly using RegTech analyses to improve the compliance process, better manage the risks, and improve efficiency. RegTech tools and techniques certainly provide a more stable and supportive foundation for navigating increasing regulations. Examples of how financial organizations utilize RegTech include:

### 6.1. Compliance Automation

RegTech tools allow financial organizations to automate compliance [9] in order to reduce staff demands for owning and managing the process. With cutting-edge technologies such as artificial intelligence and machine learning algorithms, compliance reporting and data collection and analysis can be automated. It provides real-time monitoring and short-term alert should the regulators recently change the policies. Compliance automation makes it easier for companies to identify upcoming changes and mutations in the policies, which could be explained by changes in compliance risk.

### 6.2. Risk Management

RegTech technologies are a key risk management driver [10] for financial organizations. New IT data and risk processing allow better and more accurate processing of patterns, potential risk detection. With RegTech, financial institutions are better able to detect and monitor suspicious activities and reduce compliance risks associated with money laundering. Compliance with AML regulations helps mitigate reputational damage and fines that can undermine the calibre of business operations.

### 6.3. Enhanced Data Security

RegTech helps financial institutions to secure their data, ensuring that sensitive information is well protected. This is through the deployment of encryption techniques that mediates customer data, transaction records, and other valuable data. When institutions use Robust RegTech solutions, they are fully compliant with data protection legislation, thus reassuring customers' trust.

### 6.4. Efficient Regulatory Reporting

Many Financial bodies find reporting for compliance to be tedious, a process that eats up their time. RegTech simplifies this process as it automates data extraction, analysis, and reporting. Some tools such as AI, BI, and machine learning ease the presentation of suspicious actions and report. This significantly reduces the high level of error that comes with manual reporting and saves on expenditures associated with reporting.

### 6.5. Streamlined Audit Processes

RegTech tools ease the auditing process by making data accessible. These tools generate the entire record thus giving detailed information on compliance level and progress. This eliminates the need to gather information from various corners of the institution, thus speeding the auditing process.

In conclusion, Fintech must widen its scope to incorporate RegTech that deals with regulatory compliance. This software has various cases that improve report accurateness, risk management, and overall operational efficiency of a financial institution. The software's automation, analytics, and security measure allow the institution to become more effective at navigating the complexities surrounding finance and instrumental regulatory requirements.

# 7. CONSIDERATIONS FOR CHOOSING REGTECH SOLUTIONS

When choosing the most suitable RegTech solutions for an organization, several factors need to be taken into account. The ideal solution must meet requirements as much as possible and help achieve compliance and risk management objectives. Consider the following factors:

## 7.1. Solution's Scalability

Since business expands, it is critical that the RegTech solution expands with it. Choose a solution that provides flexibility and can deal with more extensive data amounts and new regulations. This way, compliance procedures can keep up with evolving regulatory landscapes, without the need to dismantle or rebuild the system.

## 7.2. Integration Capabilities

The RegTech solution should integrate into existing systems and workflow seamlessly. Determine whether the solution is compatible with organization's infrastructure, including data management systems. Integration should cause minimal disruption to operations and allows to easily share, analyse, and report data. For instance, data gathered from disparate sources could be consolidated and analysed swiftly with RPA-enhanced compliance systems.

## 7.3. Data Security

Due to the sensitive nature of regulatory data, strong data security protections are a priority. Therefore, assess the supplied security tools of the RegTech solution, such as encryption, group role customization, and data storage restrictions. It is essential that the RegTech solution is compliant with industry standards and procedures to guarantee that confidential information is not accessed without proper authorization.

## 7.4. Regulatory Compliance

RegTech solutions were developed to help companies satisfy requisite compliance requirements. For this reason, it is critical to check if the solution corresponds with business's specific regulatory environment, such as anti-money laundering (AML) legislation or data protection laws. The all-inclusive RegTech solution must include unique compliance features tailored to industry's demands to help assure compliance and reduce the potential harm associated with non-compliance.

By critically evaluating these considerations, key decision makers in an organization will be in a better decision position to choose proper RegTech that meets compliance and risk management needs. Even so, it is important to emphasize that each company's requirements may be significantly different. As a result, it is important to customize these criteria based on firm's requirements.

The most important aspect is to identify a solution that improves the efficiency, effectiveness, and peace of mind in for compliance activities while minimizing risk. To achieve effective and efficient compliance operations and risk reduction, organizations need to prioritize scaling, integration capabilities, data security, and regulatory compliance.

## 8. FUTURE TRENDS OF REGTECH

The RegTech sector changes daily, and its future is increasingly highly tech-driven, with innovations such as AI and ML leading to a more effective compliance and risk management landscape in the financial industry.

### 8.1. Adoption of AI and Machine Learning

AI and ML algorithms are quickly becoming core components of modern RegTech solutions [5]. Through such technologies as data analytics and automation, organizations can efficiently process extensive amounts of regulatory data to ensure compliance. By integrating AI and ML, businesses can better recognize patterns, deviations, and potential risks, ensuring that they are in full compliance before each report is generated. Furthermore, powered by AI natural language processing, RegTech platforms can process unstructured data sources from regulatory reports and news articles, reaching insights to help the firm stay ahead of the curve.

### 8.2. Embracing Cloud Computing

Cloud computing redefines [2] how RegTech services are implemented and managed. In addition to investing in onsite equipment, financial firms will utilize cloud-based sites to access reliable and versatile RegTech solutions. Furthermore, cloud-based services enable collaborations with existing procedures, helping prevent catastrophe and reduce regularity. Additionally, cloud computing solutions improve data security and back-up, guaranteeing that the most significant regulatory information always is secure and accessible.

### 8.3. Focus on Data Analytics and Predictive Insights

RegTech services are also increasingly focused on powerful data analytics technologies. By investing in powerful big data analytics, a firm will have access to more knowledge about regulatory changes, compliance trends, and new regulations. Implementing forecast analytics would allow the organization to spot disruptive shifts in compliance, detect illegal patterns, and minimize potential losses. Such data analysis services inform compliance decisions and help financial firms adjust to changing regulatory markets.

### 8.4. Rise of RegTech Start-ups and Collaborations

The RegTech sector includes exciting new players developing cutting-edge technology solutions. IOT companies are designed to address frequent regulatory issues such as money laundering or fraud detection. Financial firms also team up with RegTech companies, creating incentives for their compliance regimes. Partnership between a financial company and a RegTech company helps share information quickly and generate tailor-made solutions for various industries.

### 8.5. The Increasing Importance of Regulatory Sandboxes

Regulatory sandboxes have become a critical regulatory mechanism under which regulators allow RegTech companies to test and use it for innovation without breaking the law. Financial institutions work with RegTech companies for the testing and innovation of RegTech solutions such as blockchain-based compliance management systems. This approach accelerates the use of technology to address compliance issues in the fast-changing market.

In conclusion, the future of RegTech depends on artificial intelligence, machine learning, cloud computing, data analytics, and the collaboration of financial institutions with RegTech companies. These trends have transformed compliance and risk management through the provision of the RegTech companies' service.

## 9. CONCLUSION

In conclusion, RegTech service is core in achieving the compliance of financial regulation in the financial industry's changing regulations. The use of artificial intelligence and machine learn has helped RegTech services to have new innovative means to help its customers in guarantee compliance, risk minimization, and keep, up to date with the unfolding regulation.

This paper addressed the need for the use of RegTech for compliance and risk management in the financial industry. RegTech service has a range of services that are of utmost efficacy to help in the operation and limiting or avoidance of its potential risk. Therefore, RegTech services offer financial institutions the possibility of keeping a balance between compliance and operation efficiency. RegTech services minimize the time spend and resources of the financial authorities and the monitoring of the industry for suspicious activities like fraud, risk assessment, and data collection.

Moreover, the use of RegTech solutions allow financial institutions to remain informed about any regulatory changes and combat complex hurdles of compliant. Through the use of robotic process automation, and blockchain backend system. RegTech provides a competitive advantage when it comes to dealing with the increasing difficulty of regulatory branches overhead. Selecting the correct RegTech options entails considering various aspects, including scalability, data security, integration capability, and regulation compliant.

Security concerns surrounding user data have become increasingly prevalent due to the proliferation of data breaches. Ongoing compliance is a core part of the choice for an effective RegTech platform, and any organization must confirm to comply with the most stringent standards. With regulation, the financial sector presents a dynamic field, and RegTech will continue to decentralize compliance in the future. Thus, financial institutions should employ RegTech to solve most of the organizations' problems due to the highly unregulated environments. To conclude, RegTech is not only a better choice for remaining active in the statutory market but also a needed option without any choice.

## AUTHOR

**Chintamani Bagwe** stands as a leading Fintech Expert in Banking Compliance and Risk Management, with a rich career of nearly two decades at the forefront of global banking and technology. His work centers on blending technology, data science, and business strategy to tackle the complexities of regulatory compliance. Chintamani is renowned for driving innovation and strategic development in Governance, Risk, and Compliance applications, enhancing regulatory processes, and elevating operational efficiency through his analytical prowess. He has a proven track record 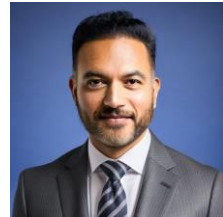 with Fortune 100 companies, managing international teams, and implementing AI-enabled financial systems across trading, risk, and collateral management. With advanced degrees in Financial Management, Business Administration, and Engineering, plus specialized certifications, Chintamani exemplifies transformative leadership in the fintech sector.